*Communication*

# Physiological Imaging: When the Pixel Size Matters

## Gennadi Saiko [1]*

[1] Department of Physics, Toronto Metropolitan University, Toronto, Canada; gsaiko@torontomu.ca

* Correspondence: gsaiko@torontomu.ca;

**Abstract:** With the proliferation of inexpensive CMOS cameras, medical imaging experiences a noticeable influx of new technologies. While anatomical imaging is based on well-established principles of photography, physiological optical imaging is a relatively novel range of technologies that requires considering a new set of technical and physiological aspects. We discuss several factors (binning, spatial frequency sampling, and distance to the target area) indispensable to getting quantifiable and reproducible results, which are the essence of physiological imaging. We also discussed their implications for several commonly used physiological imaging modalities, including hyperspectral/multispectral imaging, fluorescence imaging, and thermography. Physiological imaging technologies are in their infancy. Thus, the proper design, which considers technical and physiological aspects, is paramount to establishing their credibility and driving clinical adoption.

## 1. Introduction

With the proliferation of smartphones, healthcare is on the verge of transformational changes. The widespread use of smartphones has the potential to decrease costs, democratize healthcare, and bring diagnostic capabilities closer to patients.

One particular aspect of these transformational changes is the emergence of multiple novel medical imaging technologies. This transformation started with applications of traditional digital imaging, sometimes called anatomical imaging in this context. For example, anatomical imaging is used in wound management to track time progress [1] and dermatological applications, e.g., skin cancer diagnostics[2].

In addition to anatomical imaging, inexpensive CMOS cameras enabled imaging modalities, which extract underlying physiological information. This group of technologies can be called physiological imaging based on this intent. An example of such modalities can be photoplethysmographic (PPG) imaging for extraction of heart rate [3], hyperspectral/ multispectral imaging for extraction of tissue blood oxygenation [ 4], or fluorescence imaging of bacterial load [5]. In addition to these mainstream applications, physiological imaging can monitor capillary grid [6] or extract pulse wave velocity [ 7].

The significant subset of these technologies (other than thermographic and fluorescence imaging) is based on tissue (skin or mucosa) reflectance. In this case, physiological imaging typically samples the signal from 1.5-2mm depth [3]. Thus, the signal can be pretty faint, and its robust collection may represent some challenges.

In addition, as all physiological imaging technologies are still in the early stages of clinical translation (experimental modalities or early clinical adopters), not a large body of evidence has yet been accumulated. Thus, data quality and reproducibility issues have not emerged yet. However, this topic may get prominence with data accumulation as it happened with other optical modalities. For example, it is known that laser Doppler flowmetry (LDF) exhibits significant spatial and temporal variations in LDF measurements [ 8].

This article aims to address these potential data collection and extraction challenges from technical and physiological points of view and identify issues that may affect data quality and reproducibility during physiological imaging. Furthermore, we will generalize approaches and observations across a broad range of physiological imaging modalities. The approach can be applied to various camera types, including CMOS and CCD. However, we will illustrate it mainly using CMOS examples. Also, we will focus on lens-based applications only. While lensless applications have been proposed recently (see, for example, [9]), they are far from wide clinical adoption. Wavelength optimization (for example, narrow band imaging, [10]) is also out of scope of the current work.

## 2. Materials and Methods

The primary purpose of virtually any physiological imaging technique is to extract and quantify a physiologically relevant parameter: heart rate, tissue blood saturation, etc. Thus, we need to establish a reference point for the analytical ability of a method. In analytical methods (including clinical laboratory tests), there are two relevant levels of signal quantification: Limit of Detection (LOD) and Limit of Quantification (LOQ) (see, for example [11]).

The limit of Detection is defined as an analyte content that can be distinguished from the blank (no analyte) with an error probability of (1-$\beta$). Here $\beta$ is the probability of false-negative results; type II error.

The limit of Quantification is defined as an analyte content that can be determined with a certain level of precision.

LOD and LOQ can be defined using the mean and variance of the signal. The mean and variance can be combined in a single number metric, Signal-to-Noise ratio (SNR). LOD and LOQ are typically defined using SNR with values of 3 and 10 for LOD and LOQ, respectively.

### 2.1. CMOS Camera Noise

To assess SNR for imaging technology, we need to identify the mean and variability of the signal. Some variability of the signal can be attributed to noises. In particular, the CMOS camera has several types of noise: shot noise, dark noise, and quantization noise.

The shot noise can be modeled by a Poisson process. In this case, the variance of the noise $\sigma^2_p$ is equal to the mean number of photons $\mu_p$. The dark noise consists of read noise and dark current. However, the dark noise can be ignored in most practical cases other than low light conditions (e.g., astronomy). The quantization noise arises from digitizing the continuous voltage signal into a digital one and can be modeled by a Gaussian distribution. The read noise, dark current, and quantization noise are set values for a particular camera and typically can be found in spec sheets.

We can take into account that sources of noise are independent and use error propagation rules to write expressions for the camera output mean and SNR:

$$\mu_y = K(\mu_d + \eta\mu_p) \tag{1a}$$

$$SNR = \frac{\eta\mu_p}{\sqrt{\sigma_d^2 + \sigma_q^2/K^2 + \eta\mu_p}} \tag{1b}$$

Here $K$ is the camera's sensitivity; $\eta$ is the sensor's quantum efficiency (wavelength dependent), $\mu$ and $\sigma$ refer to mean and standard deviations for photons (subindex p), dark noise (subindex d), quantization (subindex q), and output (subindex y).

However, for simplicity, we will ignore dark noise and quantization noise. This assumption is generally valid for all cases other than low-lit conditions. In this case, the SNR can be written as (here subscript 1 refers to the single-pixel SNR):

$$SNR_1 = \frac{\eta\mu_p}{\sqrt{\eta\mu_p}} = \sqrt{\eta\mu_p} \tag{2}$$

Based on Eq.2, several observations can be derived. First, to increase SNR, we need to increase the number of photons detected. Thus, we can either increase the density of photon flux near the detector (by changing distance and increasing illumination intensity) or increase the sensor size or integration time (e.g., spatial or temporal binning). These two approaches will be briefly discussed below.

### 2.2. Distance

Distance between the imaging system and the target area may impact the collected signal in several ways. Primarily, it affects the strength of the signal, which will be discussed further. However, it also impacts the sampling depth through spatio-angular gating [12].

To analyze the impact of the distance on the signal strength, we can consider two imaging geometries: imaging the large skin area to collect averaged data (e.g., remote PPG or rPPG) and searching for small clusters within the target area (e.g., bacterial imaging or thermography). We can also analyze it across modalities with illumination (active modalities) and passive modalities, which use ambient light (e.g., rPPG) for illumination or register emitted light (thermography).

We can split the light propagation into three components:

- Incident light, which illuminates the target area (other than thermography)
- Light propagation and reflectance within the tissue (other than thermography)
- Light from the target area to the sensor

Tissue propagation and reflection will not be impacted by the distance other than mentioned above spatio-angular gating. Thus, we can focus on the forward (imaging system -> target) and backward (target -> imaging system) light propagation.

To simplify calculations, we can make further assumptions. In particular, let's assume that the light source is close to the camera in an active modality. Thus, both distances- illumination system- target area and target area- sensor are equal to $L$.

### 2.3. Binning

Binning is the grouping of outputs collected from several cells (pixels). It is an efficient way to increase sensitivity and SNR. In CCD devices, binning can be achieved on the sensor level. For CMOS devices, binning cannot be achieved on the sensor level; however, it can be done digitally. It is less efficient than sensor-level binning (for example, it cannot reduce read noise); however, it provides significant SNR improvement anyway.

In most scenarios, binning is achieved by averaging signals over $NxN$ cells. Taking into account that sources of noise in each cell are independent, we can use error propagation rules and write

$$SNR_{NxN} = \frac{\sum_{NxN} \eta\mu_p}{\sqrt{\sum_{NxN} \eta\mu_p}} = \frac{N^2\eta\mu_p}{\sqrt{N^2\eta\mu_p}} = N\sqrt{\eta\mu_p} \tag{3}$$

Thus, $NxN$ binning allows approximately $Nx$ improvement in SNR in the CMOS camera. However, it should be noted that binning reduces the image's resolution by the factor of $N$.

### 2.4. Sampling frequency

One particular application of physiological imaging is determining the parameters of the capillary grid, which can be used for shock progression monitoring [13] or cancer transformation detection [2].

In most skin parts, the capillaries are arranged vertically (hairpins). Thus, from the surface, they can be perceived as dots. However, the contrast between these points is relatively low. Therefore, to detect them, we have to account for two factors: a) the contrast ratio associated with them needs to be detectable by the sensor (pixel size ↑), and b) the spatial sampling frequency should satisfy Nyquist–Shannon sampling theorem (pixel size ↓).

## 3. Results

In this section, we will mainly use the pixel size to annotate the size of the target area, which is sampled by a camera pixel.

### 3.1. Distance

Dependences of the sensor signal on the distance from the camera to the target area, L for the large and small area imaging for the point light source, are presented in Tables 1 and 2, respectively.

**Table 1.** Dependence of the sensor signal on the distance from the camera to the target area, *L* in the case of the large area imaging (background).

|  | Active modality | Passive modality |
|---|---|---|
| Incident light intensity at the target | $\sim 1/L^2$ | constant |
| The intensity of light on the sensor from the element of surface in the target area | $\sim 1/L^2$ | $\sim 1/L^2$ |
| The area sampled per pixel | $\sim L^2$ | $\sim L^2$ |
| Total effect | $\sim 1/L^2 * 1/L^2 * L^2 \approx 1/L^2$ | $\sim 1/L^2 * L^2 \approx L^0$ |

**Table 2.** : Dependence of the sensor signal on the distance from the camera to the target area, *L* in the case of the small area imaging.

|  | Active modality | Passive modality |
|---|---|---|
| Incident light intensity at the target | $\sim 1/L^2$ | constant |
| The intensity of light on the sensor from the element of surface in the target area | $\sim 1/L^2$ | $\sim 1/L^2$ |
| The area sampled per pixel a) Target is larger than the pixel b)Target is smaller than the pixel | $\sim L^2$ $\sim L^0$ | $\sim L^2$ $\sim L^0$ |
| Total effect a) Target is larger than the pixel b)Target is smaller than the pixel | $\sim 1/L^2 * 1/L^2 * L^2 \approx 1/L^2$ $\sim 1/L^2 * 1/L^2 * L^0 \approx 1/L^4$ | $\sim 1/L^2 * L^2 \approx L^0$ $\sim 1/L^2 * L^0 \approx 1/L^2$ |

### 3.2. Binning

As we already mentioned, the realistic physiological signal can be pretty faint. For example, the amplitude of the remote PPG signal (see Figure 1) can be on a scale of 1% of the average value (background). Thus, in this case, the useful signal is the modulation, not the background itself. To illustrate it, let's suppose that the mean of the background signal is $\mu$. If we consider its slight change in time to $\mu' = \mu - \Delta$, then the SNR for physiological modulation can be found as

$$SNR_{phys} = \frac{\eta\mu - \eta\mu'}{\sqrt{\eta\mu}} = \frac{\eta\Delta}{\sqrt{\eta\mu}} = \frac{\Delta}{\mu}SNR_{background} \qquad (4)$$

Consequently, to quantify the physiological signal, we need to meet LOQ (namely, SNR at least 10:1) for the signal on the scale $\Delta/\mu$ of the background ("blank" in analytical methods terminology). Thus, for our example using Eq.4, we can estimate it as: $10 = 0.01\sqrt{\eta\mu_p}$ which requires around $10^6$ photons per measurement.

This scenario is unattainable for many sensors. For example, the pixel's full-well capacity can be on the scale of 10,000-20,000 photons. Fortunately, there is a way to achieve it using binning. For example, for 10x10 binning, the required number of photons per cell in our example will be on a scale of $10^4$, which is more plausible. However, the specific value for binning size $N$ must be determined from the relative magnitude of the expected signal and camera parameters.

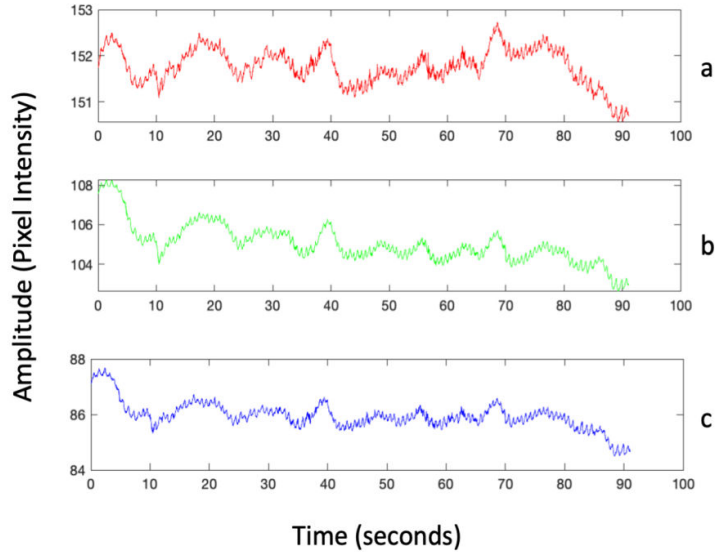

**Figure 1.** An example of a remote PPG signal. Each subplot represents raw data collected by red (a), green (b), and blue channels (c), respectively. Low-frequency (0.1 Hz) oscillations are visible in addition to the cardiac pulsations. Reproduced from [14] with permissions.

Alternatively, instead of temporal variation (PPG signal), we can consider spatial changes in reflectance (e.g., capillary visualization). In this case, $\Delta/\mu$ in Eq.4 can be substituted with the contrast ratio.

### 3.3. Sampling tissue heterogeneities

To visualize individual capillaries, we need to understand their scale. The distance between capillaries can be estimated from the mean capillary density. Tibirica et al. [15] found that in normal individuals, the mean capillary density ranges from 75mm$^{-2}$ in feet to 120mm$^{-2}$ in hands. Consequently, we can estimate intercapillary distances as $n^{-1/2}$=0.1mm.

Thus, according to the Nyquist–Shannon sampling theorem, the spatial sampling frequency should be at least two times higher than this value. Considering that this spatial sampling frequency is limited by the pixel's largest dimension (its diagonal), the maximal pixel size can be estimated at around 35μm

## 4. Discussion

Physiological imaging technologies are in their infancy. The lack of established benchmarks and complexity in comparing to the gold standard set further interpretation and clinical adoption hurdles.

Thus, the proper design, which considers technical and physiological aspects, is paramount to establishing their credibility and driving clinical adoption.

Our simple considerations show that several technical aspects can be crucial in physiological imaging data collection. Firstly, the distance is an important factor in increasing the signal's strength and reducing the necessity of binning. It is particularly important for active modalities.

When the distance and illumination intensity (for active modalities) are optimized, the binning can become essential to capture and quantify faint physiological information.

Medical imaging modalities are generally based on either still imaging (e.g., hyperspectral/multispectral imaging) or video capturing (e.g., video PPG or Pulse Wave Velocity imaging). Different techniques to improve sensitivity and SNR can be used for still imaging and video capture techniques.

For the video-capturing technique, temporal resolution is the essence of technology. Thus, spatial binning is the only way to improve SNR. However, it will decrease the resolution accordingly.

We have more flexibility for still image-capturing techniques, as temporal binning can be employed in addition to spatial binning. Multiple images can be captured and averaged to improve SNR without compromising spatial resolution. This temporal binning (image stacking) is used intensively in astrophotography.

We can briefly discuss these implications for several commonly-used physiological optical modalities.

*Remote PPG (rPPG)*. rPPG is a video-based optical modality. Thus, spatial binning is the only way to improve SNR. However, it will decrease the resolution accordingly.

*Hyperspectral/Multispectral imaging (HSI/MSI)*. HSI/MSI is usually a still image-based modality. It can be used to extract blood oxygen saturation. For this technology, spatial binning (with a reduction in spatial resolution) or temporal binning can increase SNR and sensitivity. The pathological zones (e.g., reduced perfusion) are not very small. Thus, the spatial binning on a 1mm or smaller scale should not represent an issue.

*Fluorescence Imaging*. Fluorescence imaging can be exogenous or endogenous. For exogenous fluorescence, the external dye (e.g., fluorescein or GFP) is injected into the bloodstream. In this case, blood perfusion can be visualized. Endogenous fluorescence can be used to detect bacteria or fungal presence in tissues. As it is still an image-based technique, either spatial binning (with a reduction in spatial resolution) or temporal binning can increase SNR and sensitivity. Sampling frequency can be essential to resolve blood vessels in exogenous fluorescence techniques. On the other hand, the bacterial or fungal colony size can be pretty small (less than the pixel size). In this case, distance can become significant in the endogenous fluorescence technique. In particular, the signal will be strongly ($\sim 1/L^4$) dependent on the distance, and some small targets (bacteria or fungal clusters) can be overlooked or underestimated.

*Thermography*. Thermography is a passive modality based on measuring the energy flow emitted by an object. The resolution of current bolometers is much smaller (80x60 or 160x120 for low-grade cameras) than the visible light cameras. Thus, the pixel size can be quite large (e.g., 0.5-1mm for 80x60 mm imaging area). Consequently, there can be a significant dependence on the distance for small (below pixel size) temperature anomalies. However, the body's temperature abnormities typically occur on a larger scale (1cm and above). So, in most cases, it should not be an issue.

It should be noted that with our primary focus on imaging modalities, we have considered the rectangular grid of pixels. However, the approach can be easily extended to other geometries, e.g., the circular geometry of fiber-optical probes.

## 5. Conclusions

With the proliferation of inexpensive CMOS cameras, clinical imaging experiences a noticeable influx of new technologies. While anatomical imaging is based on well-established principles of photography, optical physiological imaging is a relatively novel technology that requires considering a new set of technical and physiological aspects.

Thus, the proper design, which considers technical and physiological aspects, is paramount to establishing their credibility and driving clinical adoption.

**Funding:** This research received no external funding.

**Conflicts of Interest:** The authors declare no conflict of interest.